\documentclass[
    ,final            
  ]
  {aipproc}

\layoutstyle{6x9}

\usepackage{amsmath}
\usepackage{graphicx}
\newcommand{\V}[1]{\Vec{#1}}
\newcommand{\g}[1]{{\mathbf{#1}}}
\newcommand{\om}{\omega}
\newcommand{\bmatr}{\begin{bmatrix}}
\newcommand{\ematr}{\end{bmatrix}}
\newcommand{\cp}{\times}

\newcommand{\rV}{{\V{r}}}
\newcommand{\vV}{{\V{v}}}
\newcommand{\lV}{{\V{l}}}
\newcommand{\omV}{{\V{\om}}}
\newcommand{\Rcon}{{\cal R}}
\newcommand{\Vcon}{{\cal V}}
\newcommand{\Ucon}{{\cal U}}
\newcommand{\Mcon}{{\cal M}}
\newcommand{\RconV}{{\V\Rcon}}
\newcommand{\VconV}{{\V\Vcon}}
\newcommand{\UconV}{{\V\Ucon}}

\newcommand{\Vg}{{\g{V}}}
\newcommand{\Rg}{{\g{R}}}
\newcommand{\Hg}{{\g{H}}}
\newcommand{\HgT}{{\g{H}^\text{T}}}
\newcommand{\Mg}{{\g{M}}}
\newcommand{\U}{{\g{1}}}
\newcommand{\Fgext}{{\g{F}^\text{ext}}}

\newcommand{\Fext}{{\V{F}^\text{ext}}}

\newcommand{\beginmegj}{\null \hspace{6cm}\begin{minipage}{12.5cm}\small}
\newcommand{\emegj}{\end{minipage}\\}
\newcommand{\tnew}{\text{new}}
\newcommand{\dt}{\Delta t}

\newcommand{\gpos}{g^\text{pos}}
\newcommand{\al}{\alpha}
\newcommand{\Pt}{\partial_t}
\newcommand{\Px}{\partial_x}
\newcommand{\Q}{q}

\newcommand{\Eq}[1]{Eq.~(\ref{#1})}
\newcommand{\eq}[1]{(\ref{#1})}
\newcommand{\Fig}[1]{Fig.~\ref{#1}}

\begin{document}


\title{The contact dynamics method for granular media}

\author{Tam\'as Unger}{
  address={Department of Theoretical Physics, Budapest University of 
  Technology and Economics, H-1111 Budapest, Hungary}
}

\author{J\'anos Kert\'esz}{
  address={Department of Theoretical Physics, Budapest University of 
  Technology and Economics, H-1111 Budapest, Hungary}
}

\begin{abstract}
In this paper we review the simulation method of the \emph{non-smooth
contact dynamics}. This technique was designed to solve the unilateral
and frictional contact problem for a large number of rigid bodies and
has proved to be especially valuable in research of dense granular
materials during the last decade. We present here the basic principles
compared to other methods and the detailed description of a 3D
algorithm. We point out an artifact manifesting itself in spurious 
sound waves and discuss the applicability of the method.

\end{abstract}

\maketitle


\section{Introduction}

Granular materials, due to their rich phenomenology and wide variety of
applications in technological processes, have captured much recent interest
and are subject of active research. Their study is often based on computer
simulations, where, with growing importance, \emph{discrete element
methods} play a fundamental role. Such numerical techniques, e.g.\ the well
known soft particle \emph{molecular dynamics} \cite{cundall79,Wolf96}, the
\emph{event driven method} \cite{McNamara94,Wolf96} or the 
here presented \emph{non-smooth contact dynamics}
\cite{jean92,moreau94,Wolf96,Jean99}, have the common property that the
trajectory of each single particle is calculated by virtue of interaction
with other particles and with the environment. The characteristics of these
algorithms originate mainly from different treatments of
interparticle interactions, which leads to somewhat different
applicabilities. E.g.\ the event driven method operates with instant binary
collisions and works very well in gas-like situations, however, if a dense
state is approached, where clusters of contacting grains would appear, then
the simulation gets critically slow, a phenomenon known as \emph{inelastic
collapse}\cite{McNamara94}.

The simulation of dense granular systems is always a computational
challenge, in a typical situation dry friction and many long lasting
contacts between hard particles have to be taken into account. Using
\emph{molecular dynamics} (MD) the velocities are modeled as smooth
functions of time (even for collisions), therefore the treatment of hard
bodies becomes problematic: here small time steps of the integration must be
used resulting in slow simulation. As we will see later, this
problem of numerous hard particles is difficult to solve even with other
conceptions. In MD further difficulties are encountered by dry friction,
where it is not obvious how to distinguish between sliding and nonsliding
contacts.

The \emph{contact dynamics} method (CD), developed by M.\ Jean and J.\ J.\
Moreau \cite{jean92,moreau94,Jean99}, has a clear conception to overcome the
above mentioned difficulties. The way is to consider the particles as
perfectly rigid and to handle the interaction by means of the perfect
volume exclusion, i.e.\ the contacting bodies may not deform each
other. For that an implicit algorithm is used, which has the significant
advantage that the implementation of dry friction is rather simple; the
infinitely steep graph of Coulomb friction can be adopted, thus no
regularization is needed. The resulting method provides realistic dynamics
for various granular systems and is especially efficient for simulating
frictional multi-contact situations of hard bodies (for comparison with
experiments and with other methods see
\cite{Jean97,Radjai99,NouguierLehon2002}).

Although in recent years CD has been the simulation tool of many studies 
\cite{Radjai96,Radjai98,StaronVilotteRadjai02,Kadau02b} 
that provided understanding for several questions of granular media, 
it is rather difficult to implement this algorithm relying merely on the
literature. The reason, it seems to us, is the lack of a practical
description with hints how to adopt CD. One of the goals of this paper is
to help solving this problem: in the following section the basic ideas and
the structure of our 3D CD algorithm is presented. The description here is
confined mostly to cohesionless, rigid and spherical particles with
perfectly inelastic collisions (i.e.\ with zero Newton restitution), but it
has to be emphasized that CD contains no restriction on shape or on other
of these conditions. We refer to simulations, where cohesive disks and polygons
were implemented \cite{Kadau02b, Kadau02a}, or particles with finite Newton
restitution 
\cite{Radjai99}, even deformable bodies can be adopted in CD with a minor
change of variable \cite{Jean99}.

Finally we will discuss some consequences of the \emph{iterative solver}
used for the calculation of the interparticle forces (see later) with
respect to the computational time and we will review a potential elastic
behavior that may arise from inaccurate force-determination.

%
%

\section{The CD method}

\subsection{The basic principles}

As a \emph{discrete element method} CD provides the dynamics by
integrating the equations of motion for each particle, where besides
external fields (e.g.\ gravity) the interaction between the particles
are also taken into account. The particles are considered as perfectly
rigid and
interact with each other via point-contacts. The question is, how the force
at such a contact is calculated with CD. Because of the rigidity it cannot
be given as a function of the extent or rate of the contact-deformation,
like in the case of the soft particle MD. In fact, the principles of the
two methods are 
basically different. In CD the contact forces are calculated by virtue of
their effect, namely the generated motion has to fulfill certain
constraints. Typically such a constraint is the volume exclusion of the
particles, or the absence of sliding due to static friction.

Using constraint forces for the interaction has a serious consequence:
a contact force depends also on other forces that press the two contacting
bodies together, e.g.\ on adjacent contact forces. Thus for a compressed
cluster of rigid particles the problem cannot be solved locally for each
contact. In the algorithm, in order to calculate a global consistent system of
constraint forces at every time step, an iterative scheme is needed (called
the \emph{iterative solver}).

This iteration process of course demands computational effort, but in
exchange it makes possible to apply an implicit stepping algorithm with
large time steps. The resulting dynamics is non-smooth and includes velocity
jumps due to shocks when collisions occur. In contrast, the MD method
considers the motion smooth even for collision, therefore the harder
the particles are, the finer resolution in time is needed, consequently small
time steps are used.
\bigskip


\subsection{The constraint conditions}

\begin{figure}
\includegraphics[height=.2\textheight]{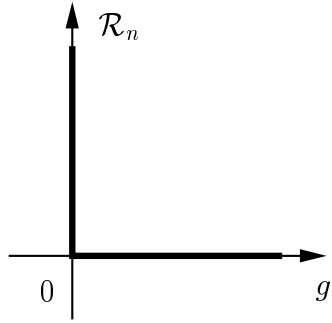}
\caption{ The Signorini graph \label{fig-signorini}}
\end{figure}

The main feature of unilateral contacts we want to attain is
impenetrability, furthermore only repulsive forces are allowed
corresponding to dry granular materials. This is expressed by the Signorini
graph (Fig. \ref{fig-signorini}), which relates the gap $g$ between two
particles and the 
normal force $\Rcon_n$ between them. $\Rcon_n$ can be non-negative and
arbitrarily large if $g=0$, but becomes zero if the two particles are not
in contact ($g > 0$). In the algorithm the normal force is determined
according to this multi-mapping graph and to the additional principle, that
the smallest value of $\Rcon_n$ is applied at a contact, which is just
needed to avoid interpenetration.

\begin{figure}
\includegraphics[height=.24\textheight]{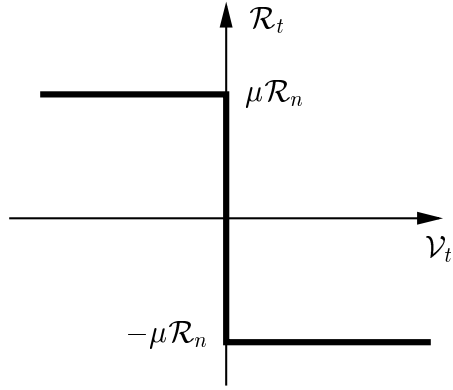}
\caption{  The Coulomb graph \label{fig-coulomb}}
\end{figure}

Regarding the tangential force $\Rcon_t$ it is due to the Coulomb friction
(Fig. \ref{fig-coulomb}), which captures the characteristics of dry
frictional contacts that 
sliding cannot be induced below a certain threshold. This threshold is
proportional to the normal force with the factor $\mu$, the friction
coefficient. (Here we neglect the difference between static and sliding
friction coefficients for simplicity, but this distinction can be made in
CD without difficulty.) 

In a non-sliding situation the value of $\Rcon_t$ is not determined by the
graph, the static friction force can be arbitrary below the threshold. Here
a similar constraint principle is applied as for $\Rcon_n$, namely that the
proper friction force is chosen, which is needed to keep the contact from
sliding. If this condition cannot be satisfied below the threshold, then
the contact will slide with a friction force $\mu \Rcon_n$ against the
motion, i.e.\ pointing opposite to the relative tangential velocity.

The graph shown in Figure \ref{fig-coulomb} concerns rather a
two-dimensional case. In 
three-dimension one can speak about the Coulomb cone: for a given $\Rcon_n$
the friction force must lie in a two-dimensional plane within a circle,
the radius of which is defined by the threshold above, thus the radius depends
linearly on the normal force.

We would like to emphasize that both graphs presented here are infinitely
steep and can be implemented in the CD method without any change.

\subsection{The discrete dynamical equations}

According to the Signorini condition, collisions of particles give rise to
shocks, therefore discontinuous velocities are expected during the 
time-evolution. In such a case of non-smooth mechanics \cite{Moreau88} the
use of 
second or higher order schemes for the integration of motion is not
beneficial and could even be problematic. Therefore first order schemes are
applied, e.g.\ an implicit Euler integration in our CD code:
\begin{eqnarray}
\rV_i\left( t+\dt \right) &=& \rV_i\left( t \right) + \vV_i\left( t+\dt
\right) \dt
\label{dx}\\
\vV_i\left( t+\dt \right) &=& \vV_i\left( t \right) +
\frac{1}{m_i}\V{F}_i\left( t+\dt \right) \dt \ .
\label{dv}
\end{eqnarray}

The two equations describe the change in velocity and position of the
center of mass during one time step for the $i$th particle. The vector
$\V{F}_i$ denotes the sum of the forces acting on this particle and is
calculated to be consistent with the new velocities and positions.

The time-stepping is also similar for the rotation around the center of
mass: the orientation of a particle is updated with the new angular
velocity $\omV_i \left( t+\dt \right)$, while for the update of $\omV_i$ we
use the torque $\V{T}_i \left( t+\dt \right)$ resulting from the new
contact forces.

\subsection{One contact}

\begin{figure}
\includegraphics[height=.2\textheight]{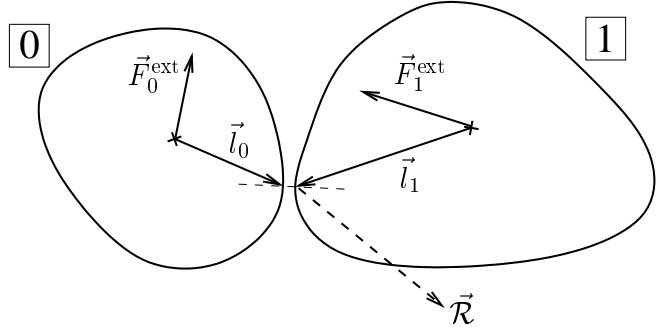}
\caption{  Two rigid particles before a possible contact
  state.\label{fig-onecontact}} 
\end{figure}

Let us first consider the simple case of only one candidate for contact,
i.e. two convex particles already in contact or with a small gap between
them. They are numbered $0$ and $1$ and may be subjected to constant
external forces $\Fext_0 , \Fext_1$ acting on the centers of mass
(Fig. \ref{fig-onecontact}). The volume 
exclusion and the Coulomb friction law may require a constraint force
$\RconV$ at this contact-candidate (contact force for short), where we
use the convention that $\RconV$ acts on particle $1$, while its reaction
force $-\RconV$ acts on the other one. In this section we will show how
$\RconV$ is calculated with \emph{contact dynamics}.

An important quantity of a contact-candidate is its relative
velocity:
\begin{equation}
\VconV=  \vV_1 + \omV_1 \cp \lV_1 - 
	\left( \vV_0 + \omV_0 \cp \lV_0 \right) \ ,
\label{contact-veloc}
\end{equation}
where
the vectors $\lV_0$ and $\lV_1$ point from the centers of mass to
possible contact points.

It is useful to introduce a generalized velocity vector:
\begin{equation}
\Vg =
\bmatr \vV_0 \\ \omV_0 \\ \vV_1 \\ \omV_1
\ematr .
\end{equation}
With the bold notation we indicate  that $\Vg$ is not a three-dimensional
vector but actually has $12$ real components. The form
(\ref{contact-veloc}) shows a linear 
dependence of the contact velocity $\VconV$ on $\Vg$.

In a similar way generalized forces can be defined:
\begin{equation}
\Rg=
\bmatr \V{R}_0 \\ \V{T}_0 \\ \V{R}_1 \\ \V{T}_1
\ematr \ \ 
\Fgext=
\bmatr \Fext_0 \\ \V{0} \\ \Fext_1 \\ \V{0}
\ematr ,
\label{RFext}
\end{equation}
where $\Rg$ contains torques and central forces equivalent to the effect of
$\RconV$:
\begin{equation}
\V{R}_0=-\RconV ,\qquad \V{R}_1=\RconV , \qquad \V{T}_0=  - \lV_0 \cp
\RconV , \qquad \V{T}_1= \lV_1 \cp \RconV ,
\label{generalized-R}
\end{equation}
while $\Fgext$ collects the external forces (the external torques are
zero).

Now the relation between the corresponding contact and central quantities
can be expressed with the following linear forms:
\begin{eqnarray}
\Rg    &=& \Hg \RconV 
\label{RgRcon}\\
\VconV &=& \HgT \Vg\ 
\label{VconVg},
\end{eqnarray}
where $\HgT$ is the transpose of the matrix $\Hg$. These two matrices
depend only on the geometry, their structures follow from
Eq. (\ref{contact-veloc}) and Eq. (\ref{generalized-R}): 
\begin{equation}
\Hg  =
\bmatr -\U \\ -\left (\lV_0 \cp \right) \\ \U \\ \left (\lV_1 \cp \right)
\ematr \ ,
\end{equation}
\begin{equation}
\HgT =
\bmatr -\U & \left (\lV_0 \cp \right) & \U & -\left (\lV_1 \cp \right)
\ematr.
\end{equation}
The elements here are $3\times 3$ matrices: $\U$ is the unit matrix and
$\left (\lV_i \cp \right)$ gives simply the crossproduct of $\lV_i$ and
the vector it is acting on. The matrix $\Hg$ allows us to transform contact
quantities into particle quantities and vice versa.

For the particles Newton's equation of motion reads:
\begin{equation}
\frac{d \Vg}{d t} = \Mg^{-1} \Rg + \Mg^{-1} \Fgext  \ ,
\label{generalized-accel}
\end{equation}
where $\Mg^{-1}$ is the inverse of the generalized mass matrix, which
is built up from the masses and moments of inertia matrices of the
particles:
\begin{equation}
\Mg =
\bmatr
m_0 \U&		0&		0&		0 \\
0&		\g{I}_0&	0&		0 \\
0&		0&		m_1 \U&		0 \\
0&		0&		0&		\g{I}_1
\ematr \ .
\label{generalized-M}
\end{equation}

If one transforms the Equation (\ref{generalized-accel}) multiplying it
with $\HgT$ from the left, the equation of motion can be obtained regarding
this candidate for contact (note that the term $d\HgT/dt \,\Vg$ describing
the geometrical change is neglected here, which is typically a good
approximation) : 
\begin{equation}
\frac{d \VconV}{d t}= \Mcon^{-1} \RconV + \frac{d \VconV^\text{free}}{d t}\
,
\label{contact-accel}
\end{equation}
where $\Mcon$ is the effective mass matrix of the contact and it
describes the contact-acceleration due to the contact
force. This is an important relation because, as we will see, it gives the
possibility to utilize our constraint conditions and therefore to calculate
the interparticle force $\RconV$.

The second term of the right hand side in Eq. (\ref{contact-accel}) has the
meaning of the 
acceleration without any interaction between the particles, i.e.\ only due
to the external forces:
\begin{equation}
\frac{d \VconV^\text{free}}{d t} = \HgT \Mg^{-1} \Fgext
\label{free-accel}
\end{equation}

\bigskip

\subsubsection{The properties of the effective mass matrix $\Mcon$}
The value of $\Mcon^{-1}$ easily follows from the transformations
(\ref{RgRcon}) and (\ref{VconVg}):
\begin{equation}
\Mcon^{-1} = \HgT \Mg^{-1} \Hg \ .
\label{invM1}
\end{equation}
It can be shown that $\Mcon^{-1}$ gets a really simple form for contacting
spheres, because on one hand the spherical moments of inertia simplify
$\Mcon^{-1}$ (i.e. $I_0$ and $I_1$ are numbers):
\begin{equation}
 \Mcon^{-1}=  \left(\frac{1}{m_0} + \frac{1}{m_1} + 
	\frac{\lV_0^{\ 2}}{I_0} + \frac{\lV_1^{\ 2}}{I_1} \right) \U
- \frac{1}{I_0}\left( \lV_0 \circ \lV_0\right)
- \frac{1}{I_1}\left( \lV_1 \circ \lV_1\right)
\label{invM2}
\end{equation} 
(where $\circ$ is the dyadic product), on the other hand one can take into
account the fact that $\lV_0$ and $\lV_1$ are on the same line, which
is perpendicular to the contact surface. The result is that $\Mcon^{-1}$
can be characterized by two parameters $m_n$ and $m_t$:
\begin{eqnarray}
m_n &=& \left( \frac{1}{m_0} + \frac{1}{m_1} \right)^{-1}
\label{m-n}\\
m_t &=& \left( \frac{1}{m_n} + \frac{\lV_0^{\ 2}}{I_0} + \frac{\lV_1^{\
    2}}{I_1} \right)^{-1} \ ,
\label{m-t}
\end{eqnarray}
which could be called normal and tangential mass respectively, as they are
the inertia of this candidate in normal and tangential
direction. More clearly if $\RconV$ is given by the sum of the normal and
tangential component: ${\Rcon}_n \V{n} + \RconV_t$ ($\V{n}$ is a normal
unit vector, while the tangential component is a vector in the two-dimensional
tangent plane), then $\Mcon^{-1}$ acts in the following way:
\begin{equation}
\Mcon^{-1}\RconV = \frac{1}{m_n} \Rcon_n \V{n} + \frac{1}{m_t} \RconV_t \ .
\label{MinvR}
\end{equation}
Thus the normal and tangential components are not coupled for spheres,
which is not true in general.

\subsubsection{The time stepping}

Turning back to the dynamics, our aim is to find a proper interaction force
in accordance with the discrete equations. Based on the implicit Euler scheme,
the following discrete form of the dynamical equations is obtained for the
two-particle system:
\begin{equation}
\Vg^\tnew - \Vg = \left( \Mg^{-1} \Rg^\tnew + \Mg^{-1}
\Fgext\right) \dt \ . 
\label{discr-generalized-accel}
\end{equation}
The superscript ``new'' indicates the yet unknown values of parameters,
i.e.\ the values after the time step $\dt$. Because the external forces are
kept constant here, 
their change is not involved in the equation. This implies the following
 approximate change in contact velocity during one time step:

\begin{equation}
\VconV^\tnew - \VconV= \left( \Mcon^{-1} \RconV^\tnew + \HgT \Mg^{-1}
\Fgext\right) 
\dt\ ,
\label{discr-contact-accel}
\end{equation}
where the unknown values are $\VconV^\tnew$ and $\RconV^\tnew$. The part of
this 
velocity change due to the external forces is known and can be subtracted from
Eq. (\ref{discr-contact-accel}):
\begin{equation}
\VconV^\tnew - \UconV= \Mcon^{-1} \RconV^\tnew \dt\ ,
\label{discr-contact-accel2}
\end{equation}
where 
\begin{equation}
\UconV =\HgT \Mg^{-1}\Fgext \dt + \VconV
\label{discr-free-veloc}
\end{equation} 
would be the new velocity
without interaction. 

Similarly, positions are updated by means of new velocities
(Eq. \ref{dx}), which implies the following predictive formula for the gap
$g$ (the distance between the two particles):
\begin{equation}
g^\tnew - g= \Vcon_n^\tnew \dt\ ,
\label{gap-change}
\end{equation}
where $\Vcon_n^\tnew= \V{n} \VconV^\tnew$ is the normal component of the
contact velocity. The normal vector $\V{n}$ points from particle $0$
towards particle $1$, thus for approaching particles the normal velocity is
negative. Note, that negative $g$ has the meaning of an overlap.

Considering the Equations (\ref{discr-contact-accel2}) and
(\ref{gap-change}) the constraint conditions can be easily imposed on the
``new'' situation, which
is done in three steps in the algorithm:

\begin{enumerate}
\item First we check what happens to the gap without interaction
  after $\dt$ and if it remains positive:
  \begin{equation}
    g+\Ucon_n \dt >0\ ,
    \label{positivegap}
  \end{equation}
  then $\RconV^\tnew$ is set to zero. This means if the contact state is not
  reached no contact force is needed. If the left hand side of the inequality
  (\ref{positivegap}) is zero or negative, the algorithm jumps to the second
  point.
\item As the two objects are considered perfectly rigid, a contact force is
  needed to hinder the interpenetration. This step is an attempt to have a
  sticking contact, i.e.\ we require on one hand that the gap closes:
  \begin{equation}
    g^\tnew=0\ ,
    \label{gapcloses}
  \end{equation}
  on the other hand no slip occurs:
  \begin{equation}
    \VconV_t^\tnew=0\ ,
    \label{stickingcondition}
  \end{equation}
  thus the unknown values can be determined. From Eq. (\ref{gap-change}) the
  new velocity is obtained $\VconV^\tnew = -(g/\dt) \V{n}$, then
  Eq. (\ref{discr-contact-accel2}) reads:
  \begin{equation}
    -\frac{1}{\dt} \Mcon \left[\left(\frac{g}{\dt} + \Ucon_n \right)\V{n}
      + \UconV_t \right] = \RconV^\tnew\ ,
    \label{noslipRnew}
  \end{equation}
  which provides the contact force. However, this contact force can only be
  accepted if it satisfies the Coulomb condition:
  \begin{equation}
    \left| \RconV_t\right| \le \mu \Rcon_n\ .
  \end{equation}
  If this inequality does not hold, the friction is not strong enough to
  ensure sticking. In this case the contact will be a slipping one and a
  new calculation is needed for $\RconV^\tnew$, which is done in the third
  point. 
\item For a slipping contact the condition (\ref{gapcloses}) remains valid,
  but (\ref{stickingcondition}) does not. Then the following equation must
  be solved:
  \begin{equation}
    -\frac{1}{\dt} \Mcon \left[\left(\frac{g}{\dt} + \Ucon_n \right)\V{n}
      + \UconV_t - \VconV_t\right] = \RconV^\tnew\ ,    
    \label{slipRnew}
  \end{equation}
  together with two further conditions corresponding to the sliding friction.
  Firstly
  \begin{equation}
    \VconV_t \text{ and } \RconV_t \text{ must be parallel and have
    opposite direction,}
  \end{equation}
  secondly
  \begin{equation}
    \left| \RconV_t\right| = \mu \Rcon_n\ .
  \end{equation}
\end{enumerate}

These three points form a \emph{shock law} that in general provides the
contact force at every time step. It can be applied for colliding
particles, but also for an old-established contact, in that sense no
distinction has to be made. If the treatment is restricted to spherical
particles, the shock law can be written in a more simple form, but
before the summary of
this case is given, we would like to make two remarks.

Firstly, this shock law corresponds to a completely inelastic collision, 
i.e.\ to zero value of the normal restitution coefficient. To accomplish such
a collision, two time steps are needed: at the first time step the relative
normal velocity is only reduced but it is not set to zero, in order to let the
gap close and at the following time step then the relative velocity
vanishes completely for the already established contact. 

Secondly, for practical applications a slight change is proposed in the shock 
law \cite{Jean99}, which is the use of
\begin{equation}
g^{\text{pos}} = \text{max}\left( g,0 \right)
\label{gpos}
\end{equation}
instead of $g$ in Equations (\ref{noslipRnew}) and (\ref{slipRnew}). This,
in principle, makes no difference because $g$ should be
non-negative. However, due to inaccurate calculations some small overlaps
can be created between neighboring particles. These overlaps would be
immediately eliminated by the first version of the inelastic shock law by
applying larger repulsive force in order to satisfy the Equation
(\ref{gapcloses}). This self-correcting property, nonetheless, has the
non-negligible drawback that it pumps kinetic energy into the system, while
pushing the overlapping particles apart, which mechanism can destroy stable
equilibrium states. With the application of $g^{\text{pos}}$ one gets rid
of the overlap correcting impulses in such a way that the already existing
overlaps are not eliminated, only the further growths are inhibited. In
that sense the resulting shock law is ``more inelastic'' than the original
one.

\medskip

The inelastic shock law of the second type is given by the following very
simple scheme for spherical particles, which allows different masses and
sizes for the contacting spheres:
\begin{equation}
\begin{array}{lrll}
\text{if} & \Ucon_n \dt + \gpos >0 & &\\
 & \text{then}&
\left\{
\begin{array}{l}
\displaystyle
\Rcon_n^\tnew:=0\\
\displaystyle
\RconV_t^\tnew:=0
\end{array}
\right.& \text{(no contact)}\\
 & \text{else}&
\left\{
\begin{array}{l}
\displaystyle
\Rcon_n^\tnew:= -\frac{1}{\dt} m_n \left(\frac{\gpos}{\dt} +\Ucon_n \right)\\
\displaystyle
\RconV_t^\tnew:= -\frac{1}{\dt} m_t \UconV_t
\end{array}
\right.
 &\text{(sticking contact)}\\
\text{if} & \left| \RconV_t^\tnew\right| > \mu \Rcon_n^\tnew & &\\
 & \text{then}&
\left\{
\displaystyle
\RconV_t^\tnew:= \mu \Rcon_n^\tnew \frac{\RconV_t^\tnew}{\left|
			    \RconV_t^\tnew\right|} 
\right.
&\text{(sliding contact)}
\end{array}
\label{sphere-shock}
\end{equation}

Note, that in the third sliding case the recalculation of $\Rcon_n$ is not
necessary, 
because the mass matrix $\Mcon$ is diagonal, even the direction of the
non-sliding friction force obtained before can be accepted for
the sliding contact, relying on the the Eq. (\ref{discr-contact-accel2}).

When situations in real experiments are numerically studied, it is a
natural requirement that certain confining objects are involved in the
simulation (e.g.\ container, fixed wall, moving piston, rotating drum).
Therefore the algorithm has to be able to handle not only sphere-sphere
contacts, but also sphere-plane and sphere-cylinder contacts. One can
easily verify that if planes and cylinders with infinite moments of inertia
are used ($I_1 = \infty$), the same simple contact law can be applied as
the one derived here for spheres. (Only in the expression of $m_t$
(Eq. \ref{m-t}) the term $\lV_1^{\ 2}/{I_1}$ is set to zero.)

\subsection{Many contacts}

As we have pointed out, the unilateral frictional problems of contacts
cannot be solved independently in a dense granular system. The unknowns of
one contact $\RconV^\tnew$ and $\VconV^\tnew$ depend on
adjacent ``new'' contact forces that are also unknown. In this way a
contact force is coupled to every other contact if they are
connected through the contact network. This is a natural consequence of the
perfect rigidity, e.g.\ a collision can immediately affect the forces
even in a very far part of the system.

In order to solve this global multi-contact problem, i.e.\  to satisfy the
constraint conditions for all contacts, an iterative method is applied in
CD. This \emph{iterative solver} is executed at every time step before the
implicit Euler integration can proceed one step further with the newly
provided forces.

This method works as follows. At each iteration step we update every
contact, independently in the sense that for one contact-candidate a 
``new'' contact force is calculated based on a one-contact shock law,
pretending that the current forces of the neighboring contacts are
constant. After that the resulting force is stored immediately as the
current force of the given contact and a new candidate is chosen for the
next update. In that way all the contact forces are updated one by one
sequentially, which of course does not yet provide a global solution, but
approaches it to some extent. Then this iteration step is repeated many 
times letting the forces relax according to their neighborhood towards a
globally consistent state. After satisfactory convergence is reached the
iteration loop is stopped. With convergence we mean that further update of
the contact forces gives only negligible changes, thus the constraint
conditions are practically fulfilled for the whole system.

 If the inelastic shock law is applied for the one-contact update, one must
not forget that the forces from the adjacent particles (acting now as
external forces) exert also torques $\V{T}_0^\text{ext}$ and
$\V{T}_1^\text{ext}$, that have to be involved by the generalized vector
$\Fgext$ in Eq. (\ref{RFext}), where the two torques originally were set to
zero.

In contrast to this sequential process a simple parallel update
would be unstable. Regarding the order of the update sequence
within the list of the candidates, it is preferably random and the random
pattern is generated repeatedly for each sweep. In this way we want to
avoid any 
bias in the information spreading, what may be caused e.g.\ by geometrical
sweeps. (If the update order is from top to the bottom, the news of a
collision or other events pass faster through the contact network downwards
than upwards, at least in the early stage of the iteration process.) It has
to be mentioned that the \emph{random sweep} described here differs from
the well known \emph{random sequential update} because while in this latter the
choice of a contact is independent from the previous choices (the same
contact could be selected even three times successively), the \emph{random
  sweep} selects each contact exactly once within one iteration step.

In applications the iteration process of the \emph{contact dynamics} has a
nice behavior and converges under reasonable assumptions. About the
conditions of convergence more can be found in \cite{Jean99} and references
therein.

\subsubsection{Convergence criteria}

Convergence criteria \cite{GuidoDiplom01} are applied in the iteration loop
to decide, whether the force-system has reached a satisfactory consistent
state or further iterations are needed. That is, a convergence criterion is
a rule according to which one choses the the final number of the
iterations within one time step. Various criteria
are used by different research groups, but the structure or efficiency of these
criteria is not much discussed in the literature.

In the following a few possible choices are presented, where the first two
examples make conditions on the relative change of the forces. The
notations used here are the following: $\RconV_\al$ is the force of the
contact-candidate $\alpha$ in the current state of the iteration process,
$\Delta \RconV_\al$ is its change during the last iteration step and the
brackets $\left< \right>$ mean average over all candidates. 

The first local criterion is:
\begin{equation}
\left| \Delta \RconV_\al \right| \le \varepsilon \left| \RconV_\al \right| \ ,
\label{localcrit}
\end{equation}
which inequality is required for every $\al$. Once it is satisfied
everywhere, the iteration loop stops. With the parameter $\varepsilon$
the required accuracy can be given. With smaller $\varepsilon$ the
calculation is more accurate, but the loop of the iteration is forced to
run longer. However, if this criterion is applied, an additional cutoff for
small forces is needed because small numerical fluctuations present in the
simulation can be relatively large for tiny contact forces (there are
always such forces in large granular packings), and therefore condition
(\ref{localcrit}) cannot be fulfilled. One way to solve this is to add a
certain small force $\delta$ to the right hand side.

The second example is of global type:
\begin{equation}
\Delta \left< \left|\RconV_\al\right| \right> \le \varepsilon \left< \left| \RconV_\al
\right| \right> \ .
\label{globalcrit}
\end{equation}
Here the change in the average force is tested, thus it means only one
condition globally for the whole system. The role of $\varepsilon$ is the
same as in the local case, but of course the same value of $\varepsilon$
provides different accuracies for the calculation of the force-system.

We performed test runs \cite{ConvCritTest} in order to compare the
efficiency of the two criteria. Here the extent of the errors and the
average iteration over many time steps needed by the simulation were
measured, this latter representing the computational effort. In most
cases the exact solution to the forces was not available, so one cannot
describe the errors as differences between exact and approximate
solutions. Here the 
level of accuracy was simply characterized by the average overlap, what
should be zero in an ideal case. This non-extensive investigation provided
the results that both criteria are working well, but in many cases
significant differences in the efficiency were found. However, which
criterion is more efficient depends on the specific situation and also on the
required level of accuracy.

\smallskip

Finally we mention the example that a criterion can simply prescribe a
fixed number of iterations: 
\begin{equation}
N_I = \text{const},
\label{constNI}
\end{equation}
i.e.\ at every time step the iteration loop stops after $N_I$ steps.
The accuracy can be improved here by
applying larger $N_I$ number. This choice of the criterion behaves also
nicely in simulations and one of its advantages is simplicity, which
makes the operations of the program more transparent giving the possibility
of a better understanding of the method.

\subsubsection{Sources of errors}

When numerical simulations are performed, numerous simplifications and
approximations are needed, which lead to dynamics somewhat different
from that in a real system. These can be e.g.\ the assumption of
point-contacts or constraint conditions when setting up a theoretical
model of a real system, or other type of approximations due to the
numerical implementation of this theoretical model, e.g.\ a discrete
integration scheme.

Knowing about the trivial sources of deviations, one can
formulate a reduced expectation from the algorithm (as we did in this
section): ``We want to have a dynamics based on the implicit Euler
integration, where the contact forces satisfy the Signorini and Coulomb
conditions.''

Is this requirement met by the here presented algorithm? In fact,
the computation of contact forces can lead to further errors. The first
and main reason of this is that the convergence of the iterative process is
not perfect, the loop is necessarily broken after finitely many steps. More
about 
the consequences of this non-accurate force calculation can be found in the
next section.  Additional errors can emerge even if the one-contact shock
law is satisfied everywhere in the system. This is because in the shock law
some geometrical changes are neglected when predicting the ``new''
configuration. E.g., for approaching particles the contact normal can
slightly turn or the vectors 
$\lV_0$, $\lV_1$ can suffer small changes, which
would alter the matrix $\Hg$.  However, if the relative displacement of the
particles during one time step is small compared to the particle size and
to the curvature of the contacting surfaces, these geometrical changes are
kept also small and their effect can be typically neglected.

\section{Spurious elastic behavior}

This section is devoted to the behavior of \emph{contact dynamics} arising
from the iterative force computation and to its consequences. The study of
the algorithm is carried out in a manner commonly applied in physics:
starting with the given microscopic laws (e.g.\ the one-contact shock law
here) properties of global behavior are examined. For a simple test
system a coarse grained description is given based on the discrete
dynamics, where physical phenomena such as 
diffusion or sound propagation will turn out to be relevant.

\subsection{Relaxation of the contact forces}

\label{forcerelax-sec}

\begin{figure}
\includegraphics[height=.12\textheight]{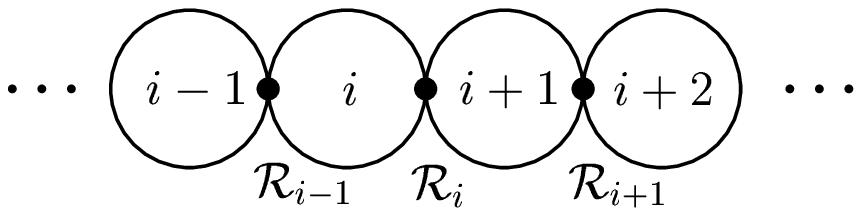}
\caption{A multi-contact situation in an 1D array. External forces
    act only on particles far away from those shown. Each particle is
    subjected only to the contact forces of the adjacent ones.%
\label{fig-chain}}
\end{figure}

In order to learn more about CD, one can analyze and test the algorithm
using the following very simple system \cite{Unger02}. The system consists of
an array of identical rigid disks or spheres aligned in a straight line
(Fig. \ref{fig-chain})
and is considered one-dimensional, as only the motion
along the line is taken into account (spinning or transversal motion is not
allowed). The long chain of particles is compressed by external forces acting
at the ends, the inner part of the chain is free from external forces. The
neighboring particles are permanently in contact, i.e.\ $\gpos_i=0$, where
the $i$th contact is between the particles $i$ and $i+1$. In this
one-dimensional case the tangential forces are zero and the normal contact
forces are denoted by $\Rcon_i$. First we show how the \emph{iterative
solver} works for this simple case.

For the update of the contact forces the shock law (\ref{sphere-shock}) is
applied. Provided the particles remain in contact and no friction is
considered, the update of the $i$th contact attains the following form:
\begin{equation}
  \Rcon_{i}:=\frac{m}{2 \dt} 
  \left(
    v_{i} - v_{i+1}\right) + 
  \frac{\Rcon_{i-1}+\Rcon_{i+1}}{2} \ .
  \label{1d-update}
\end{equation}
Based on the current values of the adjacent velocities and forces, a new
value of $\Rcon_i$ is given by this assignment which is applied several
times on the contacts in the way we described before for the
multi-contact states.  Here, the effective mass (the normal inertia of the
contacts) is half of the particle mass $m$.

As we are searching for a continuum description, the shock formula
(\ref{1d-update}) can be regarded as the discretized form of a partial
differential equation (PDE). In order to make this transformation, on one
hand, we consider the particle index $i$ as space variable $x$ and replace
the differences of consecutive quantities by spacial derivatives in the
usual way:
\begin{equation}
\frac{v_{i+1} - v_{i}}{d} \to \Px v
\label{Dxv}
\end{equation}
and 
\begin{equation}
\frac{\Rcon_{i+1}+\Rcon_{i-1}-2 \Rcon_i}{d^2} \to \Px^2 \Rcon \ ,
\label{Dx2R}
\end{equation}
where $d$ is the particle diameter.
On the other hand we introduce a fictitious time $t^*$ for the iteration
process, to be able to describe the force evolution. (Note that the physical
time does not elapse while the iterations develop the forces.) Then the
force change 
$\Delta \Rcon$ during one iteration step $\dt^*$ can be replaced by a time
derivative:
\begin{equation}
\frac{\Delta \Rcon}{\dt^*} \to \partial_{t^*} \Rcon \ .
\label{Dt*R}
\end{equation}

Thus, by applying these transformations straightforward to the assignment
(\ref{1d-update}) one arrives to the following PDE:
\begin{equation}
  \partial_{t^*} \Rcon = D \Px^2 \Rcon - \beta \Px v\\
  \label{contin-forceiter}
\end{equation}
\begin{equation}
  \text{with}\quad
  D = \Q \frac{d^2}{\dt^*}\ ,
\label{D1}
\end{equation}
\begin{equation}
  \beta = \Q \frac{md}{\dt\dt^*}\ ,
\label{beta1}
\end{equation}
\begin{equation}
  \text{and}\quad
  \Q = \frac{1}{2} \ .
  \label{q1}
\end{equation}
This analytic form clearly reveals the nature of the iteration loop:
The contact forces relax towards the solution in a diffusive way.
(The term $\Px v$ being constant during the iteration depends only on
$x$ but not on $t^*$.)

The introduction of the constant $\Q$ reflects a subtlety regarding
the sequential character of the update which was not taken into account in
the derivation of Eqs.(\ref{contin-forceiter},\ref{D1},\ref{beta1},\ref{q1}).
 In fact, $\Q=1/2$ corresponds to a parallel update, but the right
hand side of Eq.(\ref{1d-update}) always employs the freshly updated values
$\Rcon_i$, not those from the beginning of the iteration sweep. However, it can
be shown that a calculation based on the random sweep update instead of a
parallel one results in the same form of the PDE, only the value of $\Q$ is
renormalized:
\begin{equation}
  \Q=\frac{4 \sqrt{e} - 5}{2} \approx 0.797
\label{q-randsweep}
\end{equation}
(for the derivation see \cite{Unger02}). 

\smallskip

The diffusion like relaxation of the forces has an important consequence
concerning the number of iterations $N_I$ which is also crucial for
the efficiency of the algorithm, as the computational time is mainly
expended to the iterative force calculation. Due to the diffusive behavior
a long iteration process can be expected e.g.\ when an external force changes
or a collision occurs at one end of the array. One can estimate the number
of iterations required to reach an ``equilibrium'' state (convergence)
for the forces in the whole system, where the estimation is based on the
relation of two characteristic lengths: the diffusion length related to $N_I$
steps
\begin{equation}
l_\text{diff}=\sqrt{4 \,D \,\dt^* N_I}
\label{difflength}
\end{equation}
must be much larger than the system size $L$. This implies the following
estimation (q is being of order $1$)
\begin{equation}
N_I > \left( L/d \right)^2 \ ,
\label{NIestim1}
\end{equation}
that is $N_I$ scales with the square of the linear system size.

One can expect similar diffusive relaxation in dense two and
three-dimensional systems, where the diffusion takes place in a complex
contact network instead of a line, still the diffusion length is
proportional to $\sqrt{N_I}$ in typical homogen situations. Let us estimate
the total number of the contact updates during the force calculation, which
represents the computational effort of one timestep $T_\text{CDstep}$,
where the question is the scaling with respect to the number of particles
$n$. Within one sweep one update is performed for each contact, which
implies an update number proportional to $n$ (assuming the average
coordination number does not change with $n$). Therefore the computational
time of one time step is given by $n N_I$ apart from a constant factor
which results in the following scaling with the particle number:
\begin{equation}
T_\text{CDstep} \sim n^2 \quad \text{in 2D}
\label{T2D}
\end{equation}
\begin{equation}
T_\text{CDstep} \sim n^{5/3} \quad \text{in 3D}
\label{T3D}
\end{equation}
(as long as $L^2\sim n$ in 2D and $L^3\sim n$ in 3D). For comparison in soft
particle MD the computational time of one time step scales like $n$. Thus
applying the CD method in that way, it is computationally more costly for
large systems. This is the price for simulating rigid particles without
getting elasticity artifacts, which cannot be done with MD.

\medskip

\begin{figure}
    \includegraphics[height=.15\textheight]{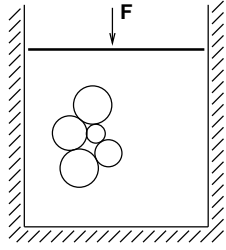}
  \caption{Setup of a numerical experiment in two dimensions. A dense
    packing of 1000 disks is prepared in a container via compressing
    the system by means of the mobile upper wall.
    \label{fig-geometry2D}}
\end{figure}

It is possible to achieve linear scaling with CD reducing the accuracy of
the force computation, however, that involves finite stiffness of the
particles. As an example let us imagine a CD simulation of a two
dimensional dense granular packing, like the one shown in \Fig{fig-geometry2D},
compressed in a container with an external force acting
upon the mobile 
upper piston. Let us suppose that the system is in equilibrium when the
piston force is increased abruptly with $\Delta F$. At the given time step
the CD algorithm starts the iterative solver to find the new force
system. What happens if one stops the iteration loop before the
convergence and performs the time step with the inaccurate forces? The
momentum $\Delta F \dt$ is transmitted from the piston through the contact
forces but only into a depth given by the diffusion length. If $N_I$ was
chosen in a way that the corresponding diffusion length is smaller than the
system height, than this momentum doesn't reach the fixed base and
consequently the upper part of the system is accelerated downwards even if
it violates the volume exclusion of the particles. We will show (based on
\cite{Unger02}) that the resulting artificial dynamics can be interpreted
as the dynamics of soft particles.

\subsection{Sound waves}

Coming back to the one dimensional system, the discrete time-evolution of
the velocities can also be transformed into a PDE in a similar manner as it was
done for the iteration loop. The Euler scheme (\ref{dv}) gives the following
update formula:
\begin{equation}
  v_i(t+\dt) := v_i(t) + \frac{\Rcon_{i-1}(t+\dt)-\Rcon_{i}(t+\dt)}{m} \dt \ ,
  \label{dv2}
\end{equation}
and changing the differences into derivatives the continuum version is:
\begin{equation}
  \Pt v = -\frac{d}{m} \Px \Rcon \ .
  \label{contin-veloc-update}
\end{equation}
This equation describes the time-evolution of the velocities on large time
and length scales. To connect it to the force update we must relate the
``iteration time'' $t^*$ to the physical time $t$. Although, depending on
the convergence criterion, there can be in principle a varying number of
iterations during one physical time step $\dt$, we assume for simplicity
that the simple criterion (\ref{constNI}) is applied, i.e.\ the number
$N_I$ is fixed.

Hence, with $\dt=N_I\dt^*$, we can express the above quantities in terms of
physical time: 
\begin{equation}
  \partial_t \Rcon = D \Px^2 \Rcon - \beta \Px v \ ,
  \label{contin-forceiter2}
\end{equation}
\begin{equation}
  \text{and}\quad
  D = \Q N_I  \frac{d^2}{\dt}
  \label{D2}
\end{equation}
\begin{equation}
  \beta = \Q N_I \frac{md}{\dt^2} \ .
\label{beta2}
\end{equation}
With the equations \eq{contin-veloc-update} and
\eq{contin-forceiter2} we obtained two coupled PDEs. We can combine
them to arrive at a wave equation with an additional
damping term:
\begin{equation}
  \Pt^2 \Rcon= c^2 \Px^2 \Rcon + \Pt\left( D \Px^2 \Rcon \right) \ .
  \label{wave-eq}
\end{equation}
The sound velocity appearing is of finite value
\begin{equation}
  c = \sqrt{\Q N_I} \frac{d}{\dt}\ .
  \label{soundvelocity}
\end{equation}

The equation~(\ref{wave-eq}) indicates that the CD simulation of the
particle chain can lead to sound propagation like in an elastic medium,
though the one-contact force update assumes perfectly rigid and completely
inelastic contacts. This deviation from the original hard particle model,
as it was mentioned before, originates from the incomplete relaxation of
the forces. As a consequence, small $N_I$ yields systematic errors in
the force calculation and involves soft particles in the sense that the
particles can overlap, furthermore, the time evolution of these overlaps
corresponds to elastic waves. The sound velocity is proportional to
$\sqrt{N_I}$, which goes to infinity in the limit of infinite $N_I$, as it
should for rigid particles.

\medskip

Searching for the dispersion relation one can perform a Fourier
transformation on \Eq{wave-eq} and 
obtain the properties of the different wave modes. The oscillation
frequency $\omega$ of the wave number $k$ is
\begin{equation}
  \omega\left(k\right)=k \sqrt{c^2-\frac{D^2 k^2}{4}} \ .
  \label{dispersion}
\end{equation}
That means, $\omega\left(k\right)$ becomes zero at a critical wave
number
\begin{equation}
  k_c = \frac{2 c}{D} \sim \frac{1}{\sqrt{N_I}}
  \label{k_c}
  ~,
\end{equation}
and waves with $k$ larger than $k_c$ (short wave lengths) are over-damped. The
damping time $\tau(k)$ for the oscillating modes is given by:
\begin{equation}
  \tau(k)= \frac{2}{D k^2} \ .
  \label{tau}
\end{equation}

We derived the dispersion relation \eq{dispersion} in the continuum
limit which is a good approximation for small wave numbers, but not
close to the border of the Brillouin zone ($k_{\text{Br}}=2\pi/d$),
where the effect of the spatial discreteness cannot be neglected. However,
increasing the number of the iterations sufficiently, $k_c$ becomes
small compared to $k_{\text{Br}}$.  Actually, for $N_I \ge 10$ the
formula \eq{dispersion} works well not only for small wave numbers but
for all oscillating modes, as can be verified numerically.

\smallskip

Related to the sound velocity and the dispersion relation one can define
the following characteristic lengths:
\begin{equation}
  l_\text{sound}= c \, \dt
  \label{lsound}
\end{equation}
is the distance the sound can travel during one time step and
\begin{equation}
  \lambda_{c}= \frac{2 \pi}{k_c}
  \label{lambdac}
\end{equation}
is the wave length of the largest overdamped mode. If $l_\text{sound}$ and
$\lambda_c$ are 
chosen much larger than the system length $L$ (with proper choice of
$N_I$), the elasticity artifacts are avoided. This is the same length-scale
as defined by the diffusion length:
\begin{equation}
l_\text{diff}=\sqrt{4 \,D \,\dt} \ ,
  \label{ldiff}
\end{equation}
since all three lengths are of the same order of magnitude: ${\cal{O}}\left(
d \sqrt{\Q N_I}\right)$. If the system is too large compared to $N_I$ in
the sense that $L > l_\text{diff}\left( N_I \right)$ the simulation
involves artificial elasticity. 

\subsection{Numerical confirmation}

\subsubsection{1D simulation}

\begin{figure}
    \includegraphics[height=.08\textheight]{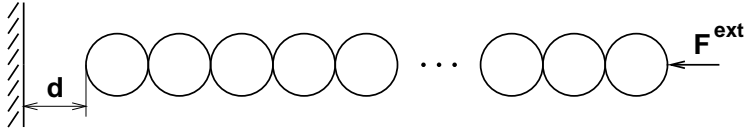}
  \caption{Initial configuration of a numerical experiment for testing
    properties of artificial sound waves.\label{fig-oscill-start}}
\end{figure}

In order to confirm the results of the coarse grained description, we
performed the following numerical experiment: The starting configuration of
the simulation consists of an array of $50$ disks and an immobile wall, the
geometry can be seen in \Fig{fig-oscill-start}. Initially the gap between
the wall and the leftmost particle is one disk diameter ($d$), the gap
between the particles is zero and the array has zero velocity.  Starting
from $t=0$ a constant external force ($F^\text{ext}$) is applied on the rightmost
particle which accelerates the array towards the wall (only horizontal
motions are present). As simulation parameters we chose $N_I=40$ and
$F^\text{ext}=0.05\:d m \dt^{-2}$.

\begin{figure}
    \includegraphics[height=.4\textheight,angle=-90]{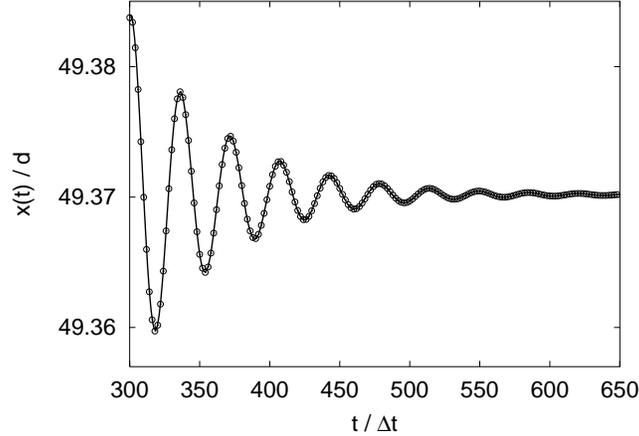}
  \caption{%
    Damped oscillation in a Contact Dynamics simulation. The dots
    indicate the measured data: the position of the rightmost particle versus
    time (for details see the text). The line is an exponentially damped sine
    function, where the frequency and the damping time is provided by the
    continuum model. \label{fig-oscill}}
\end{figure}

The collision with the wall induces a relative motion of the grains
and generates sound waves in the array. After a transient period the
grains remain permanently in contact (the whole array is pressed
against the wall by $F^\text{ext}$). Since the different wave modes have
different relaxation time, after a while only the largest wave length
mode survives. This wave length is four times the system size because
the wall represents a fixed boundary while the right side is free.
Since the wave length is given the oscillating frequency and the
damping time can be calculated from \Eq{dispersion} and \Eq{tau}. 
For comparison with the simulation we measured the
motion of the rightmost particle. The expected motion is a damped
oscillation
\begin{equation}
  x(t) = x_0+
  A \exp\left(-t/\tau\right)\sin\left(\omega t+\phi\right)
  ~,
  \label{damposc}
\end{equation}
where the parameters $x_0$, $A$ and $\phi$ are the offset, the amplitude
and the phase shift respectively. The Figure~\ref{fig-oscill} shows the
measured data (dots) and the fitted curve (\ref{damposc}) using the
calculated values of $\omega$ and $\tau$. It shows that the simulation agrees
with the coarse grained description very well.

\subsubsection{2D simulation}

After the analysis of the regular 1D system the important question
arises whether its behavior is relevant for higher dimensions and
for less regular systems. In CD simulations of
two-dimensional random packings of disks the same
``elastic'' waves can be observed (even transversal modes were found).

The simulation presented here corresponds to the thought-experiment of
the two-dimensional random dense packing we discussed before
\Fig{fig-geometry2D}. 
It consists of $1000$ disks with
radii distributed uniformly between $r_{\text{min}}$ and
$r_{\text{max}}=2 r_{\text{min}}$, the mass of each disk being
proportional to its area. 
The base and the two side-walls are fixed while the upper piston is mobile.
Starting from a loose state, we compressed the system and waited until
the packing reached an equilibrium state (the compression force $F$
applied on the piston was kept constant). The simulation was carried
out without gravity and with a Coulomb friction coefficient of $0.05$
for all the disk-disk and disk-wall contacts.

\begin{figure}
    \includegraphics[height=.4\textheight,angle=-90]{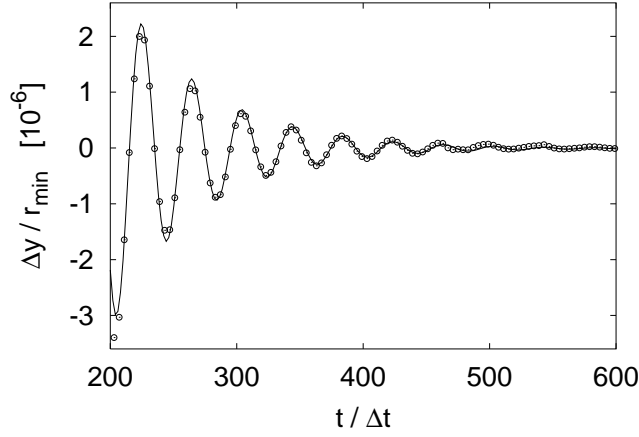}
  \caption{Oscillations in a 2D simulation are similar to the 1D case. Here the
    sound waves are generated in a random dense packing of disks. The
    dots are the measured position of the upper wall versus time (see
    \Fig{fig-geometry2D}), while the curve is a fitted exponentially
    damped sine function. \label{fig-oscill2D}}
\end{figure}

After the packing was relaxed completely, we generated sound waves by
increasing the compression force abruptly to $F+\Delta F$. After a
transient period only one standing wave mode survives (both the
wavenumber vector and the collective motion are vertical), where the
piston, representing a free boundary, oscillates with a relatively
large amplitude. We measured the vertical position of the piston versus
time and found that the data can be fitted by an
exponentially damped sine function (\Fig{fig-oscill2D}).
Here, in contrast to the 1D case, also $\omega$ and $\tau$ are fit
parameters, since, due to the different geometry, the
values \eq{dispersion} and \eq{tau} cannot be adopted. As the system has
random structure, a more complex treatment is 
required for a quantitative description. However, we checked the most
important relation, namely that the scaling properties of $\omega$ and
$\tau$ remain valid also for the 2D random system, that is $\omega
\sim \sqrt{N_I}$ and $\tau \sim N_I^{-1}$, which means that the
artificial visco-elasticity of the particles depends on the number of
iterations in the same manner as we showed for the 1D chain.

\subsection{Global elasticity}

It is instructive to compare our test-system to its simplest MD
counterpart where the contact forces depend linearly on the local
kinematic variables, i.e.\ the so called \emph{linear spring/dashpot model}
\begin{equation}
  \Rcon_i= -\kappa \left( x_{i+1}-x_i-d \right) 
       -\gamma \left( v_{i+1} - v_i \right)
  \label{lin-spring}
\end{equation}
with the spring stiffness $\kappa$ and the damping coefficient
$\gamma$. Searching again for the large scale behavior of the particle
chain one arrives to the same PDE as in
\Eq{contin-forceiter2} with its coefficients being inherited
from \Eq{lin-spring}:
\begin{equation}
  \label{spring-R}
  \partial_t \Rcon = \frac{\gamma d^2}{m} \Px^2 \Rcon - \kappa d \Px v
\end{equation}
This allows us to relate the physical MD model parameters to the
``technical'' CD parameters ($N_I$ and $\dt$):
\begin{align}
  \label{kappa}
  \kappa & = \Q m \frac{N_I}{\dt^2}
  \intertext{and}
  \label{gamma}
  \gamma & = \Q m \frac{N_I}{\dt}
\end{align}
This equivalence shows that on large scales the CD chain should behave
identical to its MD counterpart, e.g.\ it will indeed exhibit a global
shrinkage proportional to an external compressive load. Note that a
real congruence can be expected only for the collective behavior but
not on the level of contacts. In the CD method, as explained
above, contact forces are not related to the overlaps. Overlaps are present
merely due to the incompleteness of the
force-calculation and in fact are stochastic quantities because of our
random update procedure. Only on scales larger than the grain size,
where the fluctuations of these local ``deformations'' are averaged
out, the behavior can be smooth as in an elastic medium (shown e.g.\ 
in \Fig{fig-oscill}).

Performing simulations within the scope of the spurious elasticity (i.e.\
$l_\text{diff} < L$) Equations (\ref{kappa}) and (\ref{gamma}) provide
the effective stiffness and viscous dissipation of contacts. One can see
that the stiffness depends on $N_I$, but not on the total number of
particles $n$. Therefore the choice of a constant $N_I$ independently from
$n$ provides the same stiffness, no matter how large the system is. Using
the CD in that way, the superlinear scaling of the computational time in
\Eq{T2D} and \Eq{T3D} can be avoided. Thus
\begin{equation}
T_\text{CDstep} \sim n \ ,
\label{order-n}
\end{equation}
similarly to MD. In this case, of course, elasticity artifacts (sound
waves, elastic deformations) are involved by the dynamics.

 Interestingly, the reduction of $N_I$ and $\dt$ while $N_I/\dt$ is kept
constant increases the stiffness, but does not affect the value of the
dissipation $\gamma$ or the total number of updates corresponding to a
given physical time interval.
Hence, it seems to be worth reducing $N_I$ and $\dt$
in that way (at least to some degree) if the goal is to simulate harder
particles.

\medskip

The results concerning the iteration process made it possible to
characterize CD simulations by means of the diffusion length. One can
draw a distinction between two different types of behavior: spurious soft 
region for small values of $l_\text{diff}$, and rigid region for large
values of $l_\text{diff}$, where the transition can be found around the
linear system size. We would like to emphasize, however, that only dense
systems were considered so far, such states where roughly the whole system
is one large cluster of contacting particles. In loose situations, where
mainly free particles or small separate clusters are present, the system
size is not 
relevant. Here, in order to avoid the soft region, one has to compare
$l_\text{diff}$ to the size of the largest cluster.

\medskip

The description of the test system was
based on the assumption of a constant number of iterations for every
time step, and due to this premise the analytical treatment became
simple and directly comparable to the corresponding simulation. It is
important, though, that applications of other convergence 
criteria typically involve fluctuating $N_I$ (i.e.\ it can vary from time
step to time step), and therefore steps with different ``stiffness''
are mixed during the integration of motion. Consequently, the behavior
of the CD method can be more complex in detail, but qualitatively the
results of constant $N_I$ remain relevant also here. For example, the
mechanism resulting in soft particles is the same, or shock-waves with
finite velocity can also arise in the case of fluctuating $N_I$ in a
similar way. 


\section{Summary}

We presented a 3D contact dynamics algorithm in detail to simulate
large systems of rigid spherical particles and gave a review of the
iterative force calculation, where the perfect volume exclusion and the
exact Coulomb's law of dry friction are adopted. We showed that the
systematic errors can lead to a spurious collective elastic behavior and
reproduced the numerical results analytically for a simple test system.

Besides elucidating the origin of elastic behavior, the coarse grained
description revealed important characteristics of the CD
method. It was shown that using the iterative solver the contact
forces are approached in a diffusion like manner, which is a crucial
information concerning the computational time, when simulating rigid
particles properly.

When lowering the accuracy of the force calculation, CD simulation
involves soft particles. In that way Coulombian friction
can be combined with global elasticity easily and considerable
computational time can be saved: even
better performance than MD can be achieved.


\begin{theacknowledgments}

The authors thank G. Bartels, L. Brendel, D. Kadau and D. E. Wolf for valuable
discussions. Thanks are due to APS for kind permission of reproducing
figures. This work was supported by OTKA T029985, T035028.
\end{theacknowledgments}


\bibliographystyle{aipproc}   

\bibliography{granu}

\IfFileExists{\jobname.bbl}{}
 {\typeout{}
  \typeout{******************************************}
  \typeout{** Please run "bibtex \jobname" to optain}
  \typeout{** the bibliography and then re-run LaTeX}
  \typeout{** twice to fix the references!}
  \typeout{******************************************}
  \typeout{}
 }

\end{document}